\documentclass[aps,twocolumn,amsmath,amssymb,showpacs,superscriptaddress,notitlepage]{revtex4-2}
\usepackage[T1]{fontenc}
\usepackage[utf8]{inputenc}
\usepackage{color}

\makeatletter

\usepackage{amssymb}
\usepackage{amsmath}
\usepackage{graphicx}
\usepackage[colorlinks=true,linkcolor=blue,anchorcolor=red,citecolor=blue,urlcolor=blue]{hyperref}
\usepackage[caption=false]{subfig}
\usepackage{lipsum}

\makeatother

\usepackage{babel}
\def\be{\begin{equation}}
\def\ee{\end{equation}}
\begin{document}
\def\cM{{\mathcal M}}
\def\cD{{\mathcal D}}
\def\cR{{\mathcal R}}
\def\cC{{\mathcal C}}
\def\slT{\,\slash\!\!\!\! \cT}
\def\slD{\slash\!\!\!\!D}
\def\slF{\slash\!\!\!\!F}
\newcommand{\nt}{\noindent}
	
\renewcommand{\figurename}{Figure}

\title{A note on the effects of magnetic field on holographic fermions with dipole-like coupling}
\author{Sayan Chakrabarti}
\email{sayan.chakrabarti@iitg.ac.in}

\author{Debaprasad Maity}
\email{debu@iitg.ac.in}

\author{Wadbor Wahlang}
\email{wadbor@iitg.ac.in}

\address{Department of Physics, Indian Institute of Technology, Guwahati\\
Assam 781039, India}


\begin{abstract}
\nt In this note, we study the effects of non-zero magnetic field on the fermion spectral function coupled with a non-minimal coupling in the background of a dyonic $AdS_4$ black hole. The system can be reduced to a non magnetic one with the momentum being quantized into Landau levels $n$. By tuning the coupling parameter, we study its effect on the Fermi level and the quasiparticle width. We found an interesting scaling relation between  the decay width  and the coupling parameter, with a universal scaling exponent which is independent of the levels $n$.
\end{abstract}

\maketitle

\nt Despite many years of active research, strongly coupled fermionic systems still have a lot of exciting puzzles with various unconventional phenomena uncovered through condensed matter experiments like ARPES. Especially in the realm of low-temperature physics where experiments showed that though a Fermi surface still dictates the underlying features, several phenomena disagree with the conventional Landau’s Fermi
liquid theory. The so-called non-Fermi liquids \cite{Anderson:1990,Varma:1989} are an example, where a well-defined quasi-particle is absent.

Further, the problems posed by these strongly correlated systems has caught the attention of both condensed matter theorist and holographers. In recent years, the holographic method, which results from the AdS/CFT correspondence \cite{Maldacena:1997re} or gauge/gravity duality, has become an important framework to tackle these strongly coupled systems. This method is widely used to model strongly coupled dynamics in various condensed matter systems, including strange metals and high $T_c$ superconductors.\cite{Liu:2009dm,Lee:2008xf,Cubrovic:2009ye,Hartnoll:2008vx,Hartnoll:2009sz,Herzog:2009xv,Chakrabarti:2019gow}. The holographic study of the fermionic systems at a finite charge density has shown to exhibit the interesting behaviour of non-Fermi liquid using simple probe fermions in an extremal AdS black hole geometry \cite{Liu:2009dm,Faulkner:2011tm,Faulkner:2009wj}. Recently, more complicated models were also proposed, with extra symmetries included in the background gravity in order to mimic the real condensed matter system at the boundary where the observables, like the spectral function, can be extracted \cite{Andrade:2020hpu,Cremonini:2019fzz,Balm:2019dxk,Ling:2013aya,Ling:2013nxa,Andrade:2017ghg,Cremonini:2018xgj}. But some have introduce a coupling directly to the probe fermions fields in the bulk. One such coupling is the dipole coupling which has been used to study the dynamical gap in Mott physics \cite{PhysRevLett.106.091602,Vanacore:2014hka}. 

In this paper, we are going to extend such studies where a magnetic field is present. Only few works \cite{Gubankova:2010rc,Basu:2009qz,Albash:2009wz} have been done towards the study of holographic fermionic system in a non-zero magnetic field. In our study, we will explore just a tiny portion from the vastness of possibilities in this topic. We consider a non-minimal dipole-like coupling taking the dyonic AdS black hole as our background geometry. The motivation for choosing a dyonic black hole is to have both magnetic and electric charged, which correspond to a  $(2+1)$ dimensional boundary CFT with a non-zero chemical potential and a magnetic field. For this, we introduce a probe fermion with a coupling in this geometry to study the properties of the spectral functions which is the most relevant quantity directly related to condensed matter experiments.

We organise the paper as follows, in Section \ref{dbh} we discuss the set up of background geometry which is the dyonic black hole solution and further parametrize the solution parameters to arrive at the conditions for zero and non-zero temperature. In Section \ref{prfe} we discuss the fermionic action and the non-minimal coupling in the background geometry followed by solution to the Dirac equations and the spectral function. In Section \ref{resul}, we discuss and comment on our findings of the results and finally in Section \ref{conclusions} we summarize the paper with some discussion of possible future directions and final thoughts.

\section{Review of the Background geometry}\label{dbh}
\nt In order to study the affect of magnetic field on the probe fermions field, we will consider a dyonic black hole that has both electric and magnetic charges in $AdS_4$ spacetime. The most general Maxwell-Einstein action with a negative cosmological constant in 4-dimensions can be written as:
\be
S=\frac{1}{\kappa^2}\int d^4 x \sqrt{-g}\left[\cR-\frac{6}{R^2}-\frac{R^2}{g_F^{2}}F_{MN}F^{MN}\right]
\ee
Here $\cR$ is the Ricci scalar and $R$ is the radius of the AdS spacetime. From the given action, the solution to the Einstein equations in a metric form is given by 
\begin{eqnarray}
	\frac{d s^2}{R^2} &=& r^2\Big[-f(r) d t^2+d x^2+d y^2\Big]+\frac{d r^2}{r^2 f(r)} \\
	f(r) &=& 1+\frac{h^2+\mu^2}{r^4}-\frac{\left(1+(h^2+\mu^2)\right)}{r^3}\;.
\end{eqnarray}
with components of the gauge fields given by
\begin{align}
	A_y=A_r=0, \;A_t(r)= \mu(1-1/r),\; A_x(y) = -h\,y.
\end{align}
where $\mu$ is the chemical potential of the system and $h$ is the magnetic field strength, and the temperature of the system is
$T =\frac 1{4\pi }\Big(3 -  (\mu^2 + h^2)\Big).$
The condition for zero temperature is given by 
\be
h^2  + \mu ^2 =3.
\label{mag}
\ee
The relevant dimensionless  parameter related to the strength of the magnetic field is $H= h/\mu =\frac{h}{\sqrt{3- h ^2}} $ which goes from zero to infinity.
To study a system at finite temperature we set $h^2+\mu^2=3\,\eta$, where $\eta$ ranges from $0$ to $1$. The metric coefficient $f(r)$ can be written as
\be
f(r) = 1+\frac{3\,\eta}{r^4}-\frac{(1+3\,\eta)}{r^3}.
\ee
For $\eta=1$, we have the extremal black hole with temperature $T=0$ as seen above, and for $\eta=0$ we have a system at finite temperature\,.

\section{Fermion action}
\label{prfe}
\nt We will consider a bulk Dirac fermion field $\psi$ coupled to a non-minimal coupling as a probe to the system. The action for $\psi$ is given by
\begin{align}
	\label{eqn:ActionAdS4}
	S_{\psi} = \frac{1}{\kappa^2} \int d^4x
	\sqrt{-g} i\, \bar{\psi}\left(
	\,\slD\,\psi -m\,\,\psi-i\wp\,\left(\slF\,+\,\tilde{\slF}\right)\,\psi\right).
\end{align}
where \; $\slD$,\; $\slF$ and \; $\tilde{\slF}$  stands for
\begin{align}
	&\slD =e_{~a}^M\Gamma^a\Big[\partial_M
	+\frac{1}{8}\omega_M^{ab}[\Gamma_a,\Gamma_b]
	- iqA_{M}\Big],\nonumber\\&~\slF =\frac{1}{2}\Gamma^{ab}e^M_a\,e^N_b\,F_{MN},~\tilde{\slF} =\frac{1}{4}\Gamma^{cd}\epsilon_{cdab} e^a_M\,e^b_N\,F^{MN}.
\end{align}
Here $e_{~a}^M$ is the inverse vielbein,  $\omega^{ab}_{~M}$ is the spin-connection and $q$ is the fermion charge. The fermionic operator $\hat{\mathcal{O}}$ at the boundary is dual to the field $\psi$ in the bulk, and its conformal dimension $\Delta$ is mapped to spinors  $\psi$ with mass $m$  as
\begin{equation}
	\Delta = m + \frac32.
\end{equation}
\nt The Dirac equation from the above action is given by:
\begin{equation}
	(\;\slD -m-i\wp\,\slF-i\wp\,\tilde{\slF}\;)\,\psi=0
\end{equation}
which  can be written in the form
\begin{eqnarray}
	\label{unseparated}
	\Big(Z(r)+Y(y)\Big)\,\psi=0,
\end{eqnarray}
where,
\begin{align}
	&Z(r)=\sqrt{\frac{g_{ii}}{g_{rr}}}\;\Gamma^{\underline{r}}\,(\partial_r+S_p) -m\,\sqrt{g_{ii}}+\Gamma^{\underline{t}} \sqrt{\frac{g_{ii}}{-g_{tt}}}\left(\partial_t  - iq A_t\right)\nonumber \\
	&\;\;\;\;\;\;\;\;\;\;-i\wp\,\Gamma^{\underline{r}\underline{t}}\left(\partial_rA_t\sqrt{\frac{g_{ii}}{-g_{tt}g_{rr}}}+\frac{h}{\sqrt{g_{ii}}}\right)\nonumber \\ &\;\;\;\;\;\;\;\;\;\;-i\wp\Gamma^{\underline{x}\underline{y}}\left(\frac{h}{\sqrt{g_{ii}}}+\partial_rA_t\sqrt{\frac{g_{ii}}{-g_{tt}g_{rr}}}\right).\nonumber\\
	&Y(y)=\Gamma^{\underline{x}}(\partial_x - iq A_x)+\Gamma^{\underline{y}} \partial_y.
\end{align}
The term $S_p$ in $Z(r)$ is the contribution from the spin connection. If we now perform a Fourier transform along $(t,x)$, then $\partial_t\to -i\omega,~\partial_x\to ik_x$ and also we can remove the $S_p$ by rescaling the spinors $\psi=(-gg^{rr})^{-\frac14}\Psi$. Then we have $Z(r) ,\; Y(y)$ as 
\begin{align}
	Z(r)=&\sqrt{\frac{g_{ii}}{g_{rr}}}\;\Gamma^{\underline{r}}\,\partial_r-m\sqrt{g_{ii}}+\Gamma^t \sqrt{\frac{g_{ii}}{-g_{tt}}}\left(-i\omega  - iq A_t\right)\nonumber\\&-i\wp\,\Gamma^{\underline{r}\underline{t}}\,P(r)-i\wp\Gamma^{\underline{x}\underline{y}}\,P(z) \;,
	\nonumber\\
	Y(y)=&\Gamma^{\underline{x}}\cdot(i\,k_x - iq A_x)+\Gamma^{\underline{y}}\cdot \partial_y.
\end{align}
where, $P(r)=\,\frac{(h+\mu)}{r}$.
For convenience, we will choose the following Gamma matrices:
\begin{align*}
	&\Gamma^{\underline{r}}=\left( \begin{array}{ccc}1\!\!1 & 0  \\0 & -1\!\!1 \end{array}\right),\;
	\Gamma^{\underline{t}}=\left( \begin{array}{ccc}0 & i\sigma_3  \\i\sigma_3 & 0 \end{array}\right),\\
	&\Gamma^{\underline{x}}=\left( \begin{array}{ccc}0 & \sigma_1  \\ \sigma_1 & 0 \end{array}\right),\;
	\Gamma^{\underline{y}}=\left( \begin{array}{ccc}0 & \sigma_2  \\ \sigma_2 & 0 \end{array}\right)\;.
\end{align*}
Then we have
\begin{align*}
	&Z\,(r)=\left(\begin{array}{ccc}  \cD^-_r & i\sigma_3 \,V_+\\ i\sigma_3 \,V_- & -\cD^+_r
	\end{array}\right),  \\
	&Y\,(y)=\left(\begin{array}{ccc}
		0 & \sigma_1\cC_y +\sigma_2 \partial_y \\
		\sigma_1 \cC_y+\sigma_2 \partial_y & 0
	\end{array}\right).
\end{align*}
where
\begin{align}
	&\cD^{\pm}_r=\,\sqrt{\frac{g_{ii}}{g_{rr}}}\,\partial_r\,\pm m\,\sqrt{g_{ii}}\;+\wp\,P(r)\,\sigma_3\,,\;\; \cC_y=i(k_x + qhy)\,\;.\nonumber \\
	&V_{\pm}= -i\;\Bigg(\;\sqrt{\frac{g_{ii}}{-g_{tt}}}\left(\omega  +q A_t\right)\;\pm\; \wp\,P(r)\;\Bigg)\;\; 
\end{align}

\nt Clearly, the matrices $Z$ and $Y$ do not commute. However, it is possible to find a constant matrix $\cM$ such that $[\cM Z,\cM Y]=0$  and then look
for common eigenvectors of $\cM Z$ and $\cM Y$ as they are
commuting Hermitian operators.
The transformation matrix statisfies the following relations:
\begin{align*}
	&\bigl\{\cM,\Gamma^{\underline{r}}\bigr\}\,=0\;,\;\;\;\bigl\{\cM,\Gamma^{\underline{t}}\bigr\}\,=0\;,\;\;\big[\cM,\Gamma^{\underline{x}}\big]\,=0\;,\\
	&\big[\cM,\Gamma^{\underline{y}}\big]\,=0\;,\;\big[\cM,\Gamma^{\underline{x}\,\underline{y}}\big]\,=0\;,\;\;\big[\cM,\Gamma^{\underline{r}\,\underline{t}}\big]\,=0\;.
\end{align*}
It turns out that a convenience choice for $\cM$ is $ \left(\begin{array}{ccc}  0& \sigma_3 \\ -\sigma_3 &0
\end{array}\right)$.
In order to separate the variables, one can  multiply (\ref{unseparated}) by $\cM$ from the left to have
\begin{align*}
	&\cM \;Z=\left(\begin{array}{ccc}  i\,V_- & -\sigma_3 \cD^+_r \\  -\sigma_3 D^-_r & -i\, V_+\end{array}\right),\\
	&\cM\; Y=\left(\begin{array}{ccc}
		\sigma_3(\sigma_1\cC_y+\sigma_2  \partial_y) & 0 \\
		0 & -\sigma_3(\sigma_1\cC_y+\sigma_2  \partial_y)
	\end{array}\right).
\end{align*}
Now we can look for solutions of the eigenvalue equation of the form
\begin{eqnarray}
	\cM Z\Psi=-\cM Y \Psi=\lambda \Psi.
\end{eqnarray}
with real $\lambda$. Writing $\Psi\,=\,(\Psi_+\;,\Psi_-)^T$.
First we solve the $y$ dependent part of the above equation, which is identical to that of a massless free fermion in ($2+1$) dimensions
\[
\cM Y\Psi=\left( \begin{array}{ccc}\sigma_3(\sigma_1\cC_y+\sigma_2  \partial_y) \Psi_+ \\
	-\sigma_3(\sigma_1\cC_y+\sigma_2  \partial_y) \Psi_-  \end{array}\right)
=-\lambda\left(\begin{array}{ccc} \Psi_+ \\ \Psi_-\end{array}\right).
\]
or,
\[U_1\Psi_+\,=\,-\lambda\,\Psi_+\;\text{and}\;U_2\Psi_-\,=\,-\lambda\,\Psi_-\]
where $U_{1,2}\,=\,\pm \sigma_3(\sigma_1\cC_y+\sigma_2  \partial_y) $.
For the $\Psi_+$ part, let us left multiply $U_1$, then
\[U_1\,U_1\Psi_+\,=-\,U_1\,\lambda\,\Psi_+\;=\,-\lambda\,U_1\Psi_+\,=\,-\lambda(-\lambda\Psi_+)\]
\begin{eqnarray}
	\left[\sigma_3(\sigma_1\cC_y+\sigma_2  \partial_y)\right]\left[\sigma_3(\sigma_1\cC_y+\sigma_2  \partial_y)\right]\Psi_+=\lambda^2 \Psi_+.
\end{eqnarray}
On further simplification (16) gives
\begin{eqnarray}\label{eq17}
	\partial_y^2\;\Psi_+  + (\lambda^2+C_y^2-i\sigma_3 C'_{y})\Psi_+=0.
\end{eqnarray}
Now separating the variables $r$ and $y$ as  $\Psi_+=\Big(Z_1(r)Y_1(y),Z_2(r)Y_2(y)\Big)^T$, and noting  equation (\ref{eq17}) does not depend on $r$, one can set $Z_1=Z_2$, leading to
\begin{eqnarray}
	-\partial_y^2\; Y_{1,2} +T_{\pm}(y)\;Y_{1,2}=0,
\end{eqnarray}
where $T_{\pm}(y)=-(\lambda^2+C_y^2\mp i C_y')=(k_x + qh\,y)^2-\lambda^2 \mp qh$,
which is similar to a simple harmonic oscillator differential equation. To see this more explicitly, let us define
$\tilde{x}=\sqrt{q h}\;(y + \frac{k_x}{qh})$. Then we have
\begin{eqnarray}
	-\partial_{\tilde{x}}^2\; Y_{1,2}+\tilde{x}^2 Y_{1,2}=\left(\frac{\lambda^2}{ qh} \pm 1\right)Y_{1,2}.
\end{eqnarray}
These equations do not have the form of the standard Hermite differential equation that arises in the case of harmonic oscillator problem. But on rescaling $Y_{1,2}=e^{-\frac{\tilde{x}^2}{2}}\;Y_{1,2}(\tilde{x})$, we have
\begin{eqnarray}
	\partial_{\tilde{x}}^2\; Y_{1,2}-\tilde{x}\,\partial_{\tilde{x}} \;Y_{1,2}+(E^{\pm}-\,1)\;. Y_{1,2}=0\nonumber
\end{eqnarray}
Therefore by setting $E^{\pm}-1\,=\,2n$,
where $n$ denotes the Landau levels, the eigenvalues and eigenvectors are given by
\begin{eqnarray}
	E_n &=& \frac{1}{2}\left(\frac{\lambda^2}{qh} \pm 1\right)=n+\frac{1}{2}, ~~n=0,1,2,\dots  \nonumber \\
	Y_{1,2} &=& N_n e^{-\tilde{x}^2/2} H_n(\tilde{x})\equiv I_n(\tilde{x}),
\end{eqnarray}
where $N_n$ is a normalization constant. By substituting this solution back into the first order equations for $\Psi_+$ assuming the same eigenvalues we get
\[ \Psi_+=\left(\begin{array}{ccc} I_{n}(\tilde{x})~Z_1 \\ -i I_{n-1}(\tilde{x})~Z_1 \end{array}\right),\]
whose eigenvalues are $\lambda_n= \sqrt{2nqh},
~n=0,1,2,\cdots$. For completeness, we define $I_{-1}(\tilde{x})=0$.
Together with the expression for $\Psi_-$, which can be obtained in a similar fashion, we have
\begin{eqnarray}\label{gen1}
	\Psi=\left(\begin{array}{ccc} I_{n}(\tilde{x})~Z_1 \\ -i I_{n-1}(\tilde{x})~Z_1 \\ I_{n}(\tilde{x})~Z_2 \\ i I_{n-1}(\tilde{x}) ~Z_2\end{array}\right).
\end{eqnarray}

\nt Another independent solution comes from the values corresponding to $-\lambda_n$, i.e.
\begin{eqnarray}
	\cM\,Z\,\tilde{\Psi}=-\cM\,Y\,\tilde{\Psi}=-\lambda_n\tilde{\Psi}\;.
\end{eqnarray}
Proceeding in a similar way, we get
\begin{align}\label{gen2}
	\tilde{\Psi}=\left(\begin{array}{ccc}
		-i~I_{n}~\tilde{Z}_1 \\  I_{n-1}~\tilde{Z}_1 \\ -i~I_{n}~\tilde{Z}_2 \\ - I_{n-1}~ \tilde{Z}_2
	\end{array}\right).
\end{align}

\nt The equations for $Z_i$ and $\bar{Z}_i$ are respectively,
\begin{align}\label{foureqn}
	&&\Bigg(\sqrt{\frac{g_{ii}}{g_{rr}}}\partial_r\,-m\,\sqrt{g_{ii}}\pm\wp\,P(r)\Bigg)Z_1=-(i\,V_+\,+\lambda_n)Z_2\;,\nonumber\\ &&\Bigg(\sqrt{\frac{g_{ii}}{g_{rr}}}\partial_r\,+m\,\sqrt{g_{ii}}\pm\wp\,P(r)\Bigg)Z_2=\;(i\,V_-\,-\lambda_n)Z_1 \;,\nonumber\\
	&&\Bigg(\sqrt{\frac{g_{ii}}{g_{rr}}}\partial_r\,-m\,\sqrt{g_{ii}}\pm\wp\,P(r)\Bigg)\tilde{Z}_1=-(i\,V_+\,-\lambda_n)\tilde{Z}_2\;,\nonumber\\ &&\Bigg(\sqrt{\frac{g_{ii}}{g_{rr}}}\partial_r\,+m\,\sqrt{g_{ii}}\pm\wp\,P(r)\Bigg)\tilde{Z}_2=\;(i\,V_-\,+\lambda_n)\tilde{Z}_1 \;.
\end{align}

\nt To derive the flow equation and the Green's function, we follow the construction from \cite{Basu:2009qz,Gubankova:2010rc}, by writing the general solution as linear combination of both (\ref{gen1}) for $\lambda_n$ and (\ref{gen2}) for ($-\lambda_n$) as follows
\begin{align}
	\Psi&=\sum_n\Psi^{(n)}, \;\;	\Psi^{(n)}=\Psi^{(\lambda_n)}+\Psi'^{(-\lambda_n)}\nonumber\\
	&=\left(\begin{array}{ccc} iZ_1^{(n)}e_1^{(n)} \\  Z_2^{(n)}e_2^{(n)} \end{array}\right)
	+\left(\begin{array}{ccc} -i\bar{Z}_1^{(n)}e_2^{(n)} \\  -\bar{Z}_2^{(n)}e_1^{(n)} \end{array}\right),
\end{align}
where, $e_1^{(n)}$ and $e_2^{(n)}$ are the basis spinors. In terms of the above Hermite polynomials, the basis spinors are given by 
\[
e_1^{(n)}=\left(\begin{array}{ccc} i I_{n}(\tilde{x}) \\  I_{n-1}(\tilde{x}) \end{array}\right),\;\;
e_2^{(n)}=\left(\begin{array}{ccc}  I_{n}(\tilde{x}) \\ i I_{n-1}(\tilde{x}) \end{array}\right), \qquad
n\geq 1.
\]
Using the orthonormality of Hermite polynomials from the $y$ dependent equations one can construct the retarded Green's function which is given by \cite{Liu:2009dm,Basu:2009qz}

\begin{align}
G_R^{(n)}=-i S^{(n)}\gamma^t 
=\left(\begin{array}{ccc}  -\frac{Z_2^{(n)}}{Z_1^{(n)}} &0\\ 0& \frac{\bar{Z}_2^{(n)}}{\bar{Z}_1^{(n)}} \end{array}\right).
\end{align}
Then the spectral function $A$ is
\begin{eqnarray}
	A^{(n)}
	\sim  {\mathrm{Im}}\left(-\frac{Z^{(n)}_2}{Z^{(n)}_1}+\frac{\bar{Z}^{(n)}_2}{\bar{Z}^{(n)}_1}\right).
	\label{spect}
\end{eqnarray}

\nt By defining $G_{11}^{(n)}=\frac{Z_2^{(n)}}{Z_1^{(n)}},G_{22}^{(n)}=\frac{\bar{Z}_2^{(n)}}{\bar{Z}_1^{(n)}}$, and combining with equation (\ref{foureqn}), we get the following flow equations
\begin{align}
	&\sqrt{\frac{g_{ii}}{g_{rr}}}\partial_r\,G_{11}^{(n)} = -2m\,\sqrt{g_{ii}}\,G_{11}^{(n)}+(i\,V_--\lambda_n)\nonumber\\&\hspace{3cm}+(i\,V_++\lambda_n)(G_{11}^{(n)})^2\;,\nonumber \\
	&\sqrt{\frac{g_{ii}}{g_{rr}}}\partial_r\,G_{22}^{(n)} = -2m\,\sqrt{g_{ii}}\,G_{22}^{(n)}+(i\,V_-+\lambda_n)\nonumber\\&\hspace{3cm}+(i\,V_+-\lambda_n)(G_{22}^{(n)})^2\,.
	\label{maineq}
\end{align}
These two equations are exactly the same equations appearing in equation (14) of \cite{Vanacore:2014hka} with $G_{11}^{(n)}\sim \xi_+,\;G_{22}^{(n)}\sim \xi_-$ , $\lambda_n\sim k$. Surprisingly, the magnetic field $h$ from the dual $\tilde{\slF}$ does not appear in the flow equations explicitly. We will analyse both the Green's function and the spectral function, which is given by
\be
A^{n}(\omega,\lambda_n)\,=\,\text{Im}\,(G_{11}^{(n)}+G_{22}^{(n)}).
\ee

\section{Results and Discussion}\label{resul}
\nt In this section we will mainly focus on the magnetic field's effect and the effect of the dipole coupling on the Fermi level and the quasiparticle's nature. At this stage, we would like to mention a
few observations  before studying the properties of the spectral function  with the effects of a dipole parameter $\wp$. Firstly, we noticed from the flow equation (\ref{maineq}), the role of momentum $k$ discussed in \cite{Liu:2009dm} is now replaced by $\lambda_n$ which is discrete because of the levels $n$. For zero temperature analysis, it is convenient to write the blackhole charge $Q=\sqrt{3- h ^2}$ in terms of $h$, and then define the effective charge of the system as $q_{s}=q_0\sqrt{1-\frac{h^2}{3}}$, such that $q_0$ will be the total charge in the absence of a magnetic field. One can map back to a non magnetic system by taking the limit $h\rightarrow 0$ and $n\rightarrow \infty$, and plotting Im$G_{22}$ as a function of $\omega$ shown in Figure \ref{fig0h} below. This plot approximately matches with the results of \cite{Liu:2009dm}.\par 
\begin{figure}[htbp]
	\begin{center}
		\includegraphics[width=0.7\columnwidth]{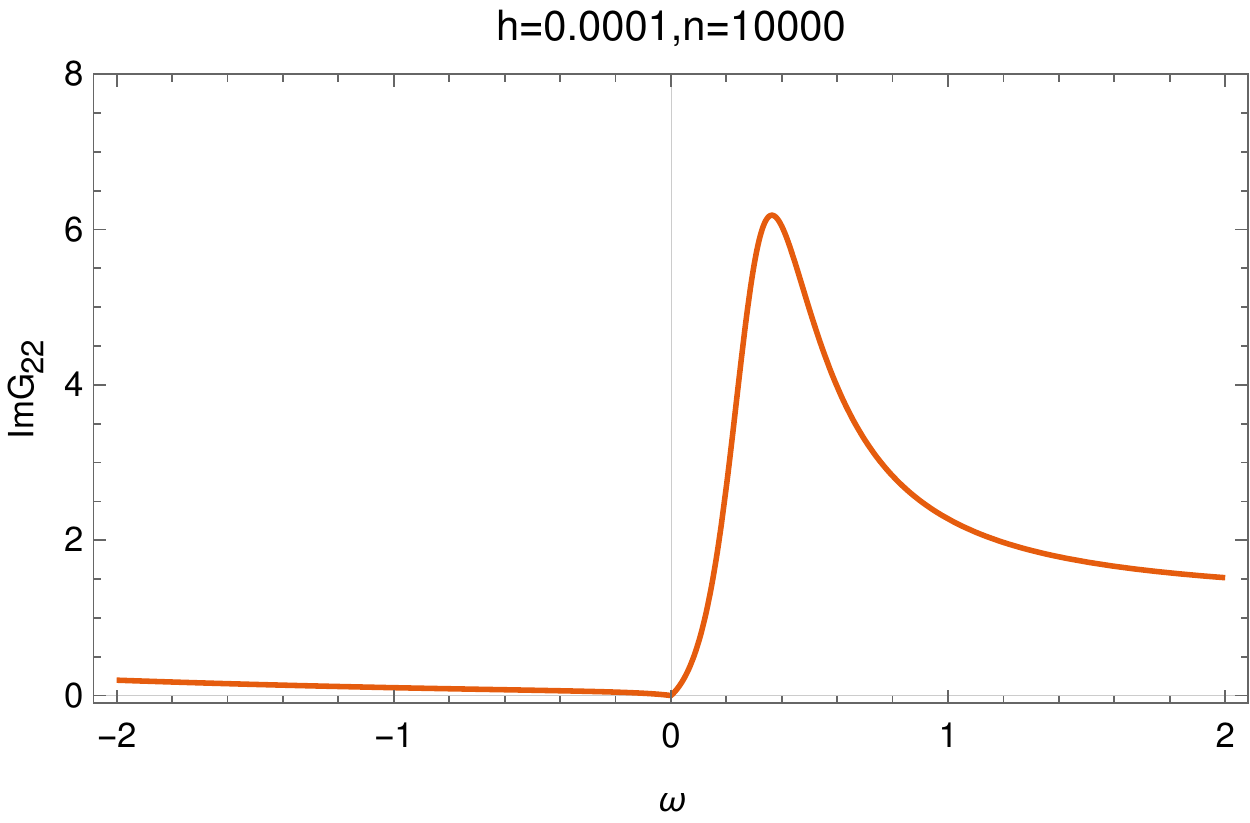}
		\caption{Zero temperature Im\,G$_{22}$ for $m=0$ and $q_{0}=1,\wp=0$. In the limit of $h\rightarrow0$ and $n\rightarrow\,\infty$.}
		\label{fig0h}
	\end{center}
\end{figure}

\subsection{Effects of the dipole coupling}
\begin{figure}[h]
	\begin{center}
		\includegraphics[width=\columnwidth]{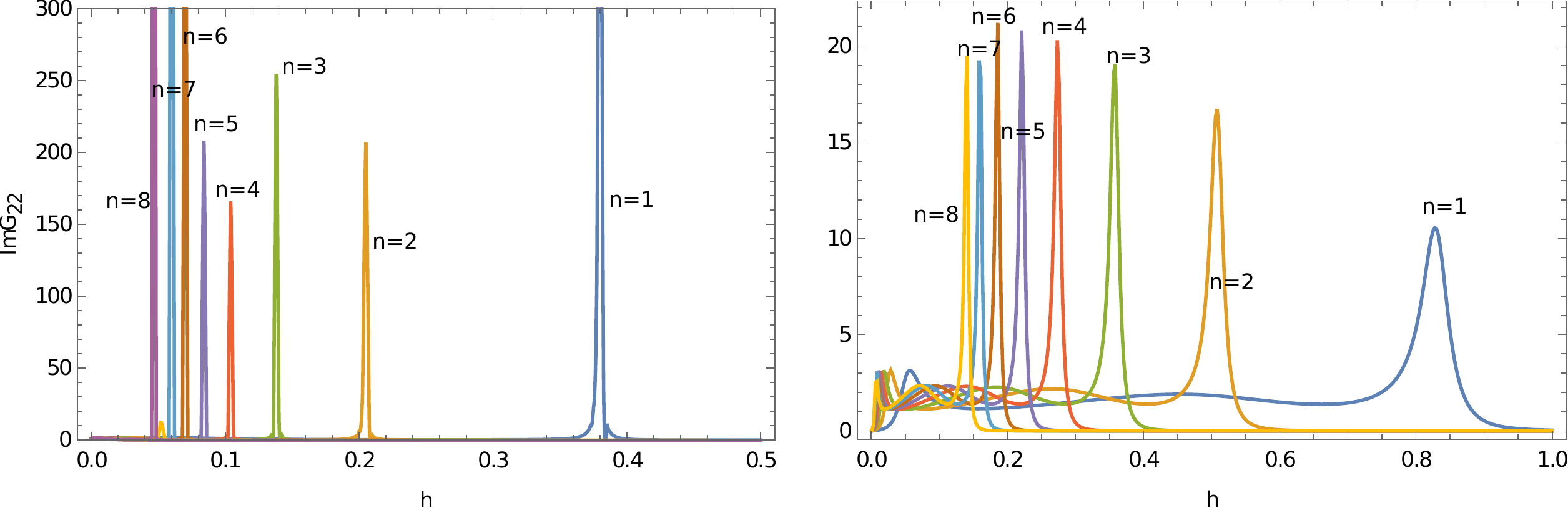}
		\caption{Left: Plot of Im\,G$_{22}$ for zero temperature, the poles  near $\omega=0$ is clearly visible for $n=1,2,3...$ levels at different field strength $h$ with $m=0$ and $q_{0}=1,\wp=0$. Right: Here, $\wp=0.5$ and the height of quasi particle decreases.
		}
		\label{fig0}
	\end{center}
\end{figure}

\nt Since, the effective momentum $\lambda_n$ is discrete depending on the level $n$,  before taking into account the detailed contribution from the constant coupling parameter $\wp$, we discuss the case with $\wp=0$. Towards that direction we first look at the behaviour of Im$G_{22}$ as function of magnetic field $h$ at zero temperature. Figure \ref{fig0} (left panel), shows the discreteness in the spectrum, which  are unique in the presence of a magnetic field. For every discrete $n$ values, there is a corresponding pole, which peaks whenever the effective momentum $k_{\text{eff}}=\sqrt{2qnh}$ equals the Fermi momentum $k_F$. Also for larger $n$, the poles are moving towards smaller $h$ and the spacing between these poles decreases. Other interesting features without the coupling terms were studied in details by the authors in \cite{Albash:2009wz,Gubankova:2010rc,Basu:2009qz}. However we are interested for the case when our coupling is non-zero. Though, extensive studies have been done for holographic fermions with dipole coupling for non dyonic background, here we generalize the discussion for a dyonic black hole. As a quick preview, from Figure \ref{fig0} (right panel), we see that by turning on the dipole parameter $\wp=0.5$, the pole's height reduces significantly even for the same magnetic field strength keeping the discrete nature of Im\,G$_{22}$ intact.

\begin{figure}[h]
	\begin{center}
		\includegraphics[width=0.75\columnwidth]{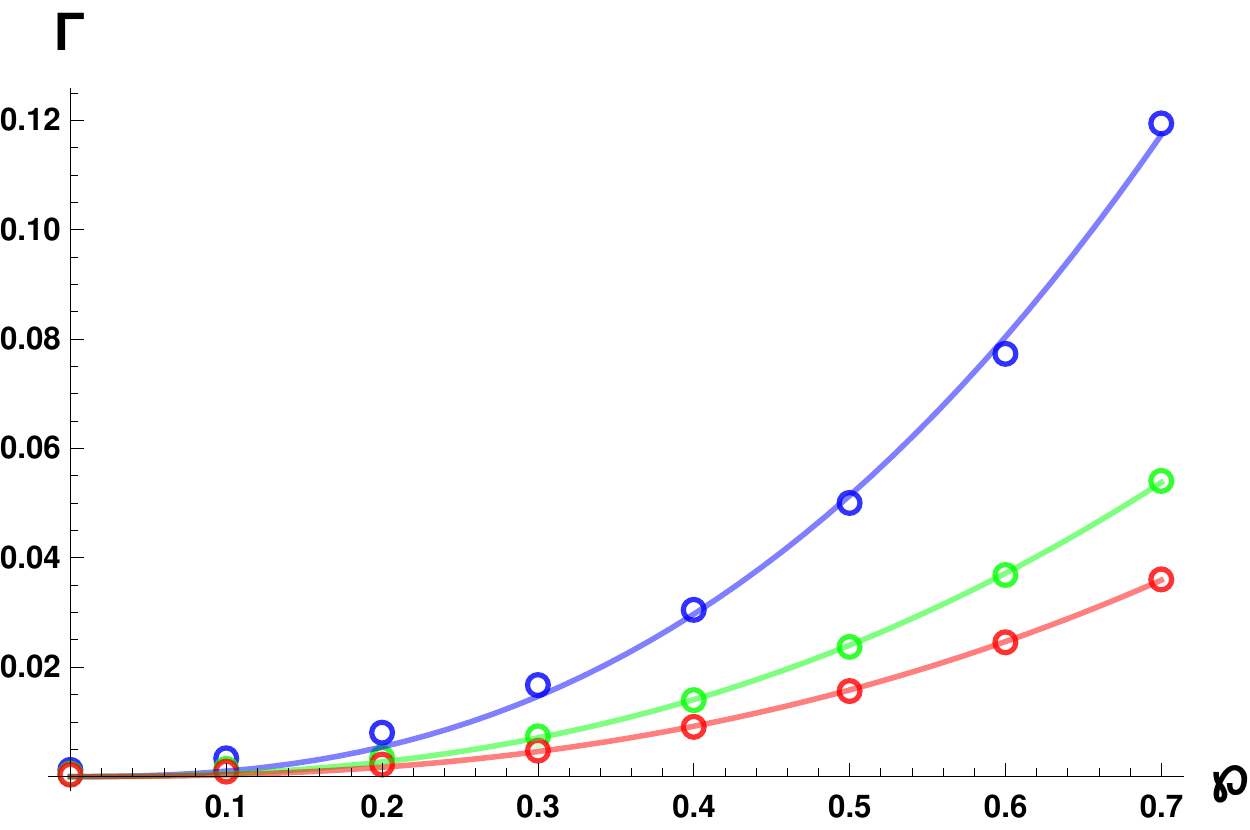}
		\caption{ Variation of quasiparticle's decay width $\Gamma$ (lifetime $\tau\sim\Gamma^{-1}$) as a function of $\wp$ for $n=1$(blue), $n=2$(green) and $n=3$(red) near $\omega=0$. The open circular markers correspond to the decay width data, while the solid line are the fitted curves in (\ref{fittedcurve}) with $\alpha\approx(0.28,0.12,0.08)$. Here, fermion mass $m=0$, charge $q_0=1$.
		}
		\label{fig02}
	\end{center}
\end{figure}

\nt Further, taking a particular level $n$, we study the decay width $\Gamma$ of a quasiparticle, which in turn gives the life-time $\tau$ by a relation $\tau\sim \Gamma^{-1}$. In Figure \ref{fig02}, we showed the variation of the decay width versus the coupling constant $\wp$ for the levels $n=1,2$ and $3$. The plot indicates that the width $\Gamma$ is increasing as we increase $\wp$. This means that the quasiparticle lifetime decreases with increasing $\wp$. By performing a non linear fit on the decay width data obtained from Figure \ref{fig02}, we get a relationship between the width and $\wp$ as 
\begin{align}\label{fittedcurve}
	\Gamma=\,\alpha(n)\,\wp^{\frac{12}{5}}
\end{align}
where $\alpha$ is a constant, which depends on $n$. The numerical values of $\alpha(n)$ corresponding to $n=1,2$ and $2$ plots are given in the caption of Figure \ref{fig02}. The most important feature of equation (\ref{fittedcurve}) is the universality of the exponent for all the $n$ values. 
\begin{figure}[htbp]
	\begin{center}
		\includegraphics[width=0.75\columnwidth]{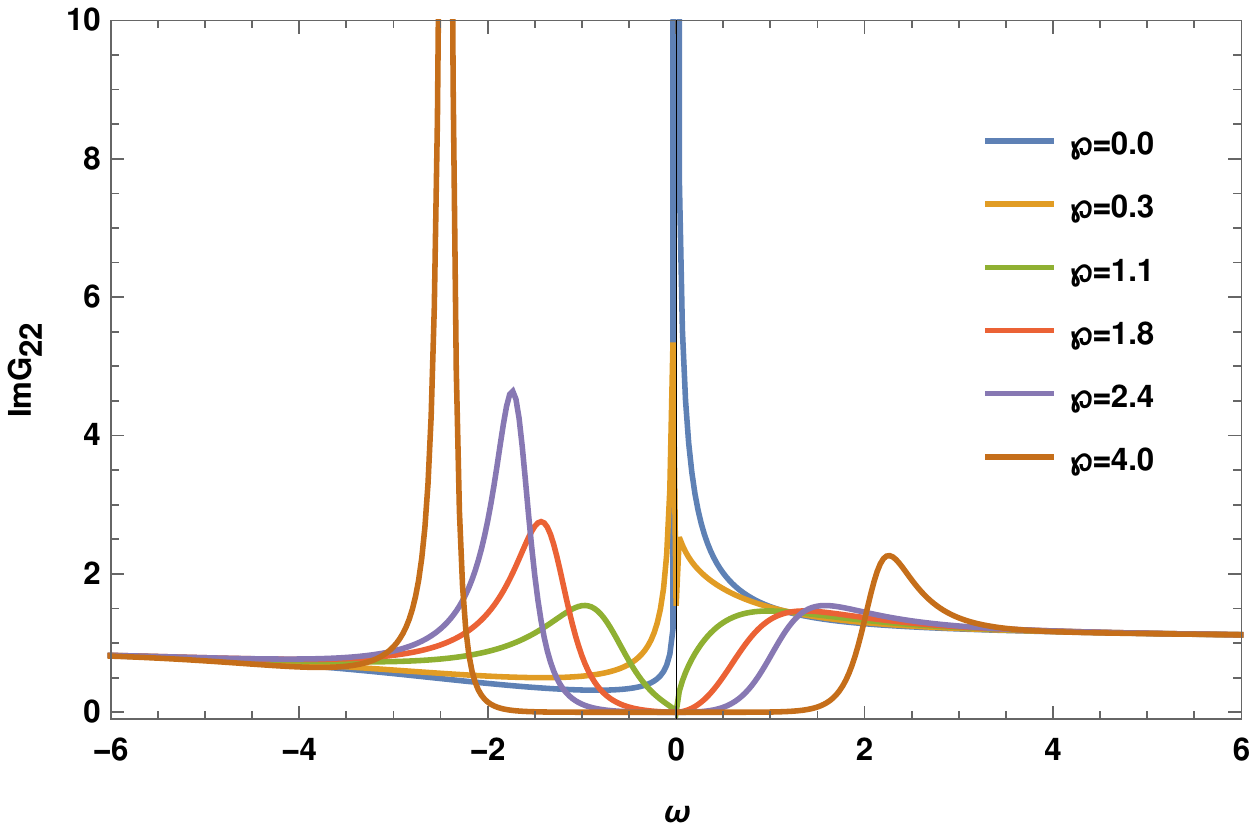}
		\caption{Gap induced by the coupling $\wp$  near $\omega=0$. Here, fermion mass $m=0$, charge $q_0=1$, magnetic field $h=0.38$ and level $n=1$.
		}
		\label{fig01}
	\end{center}
\end{figure}
%
\begin{figure}[htbp]
	\begin{center}
		\includegraphics[width=0.75\columnwidth]{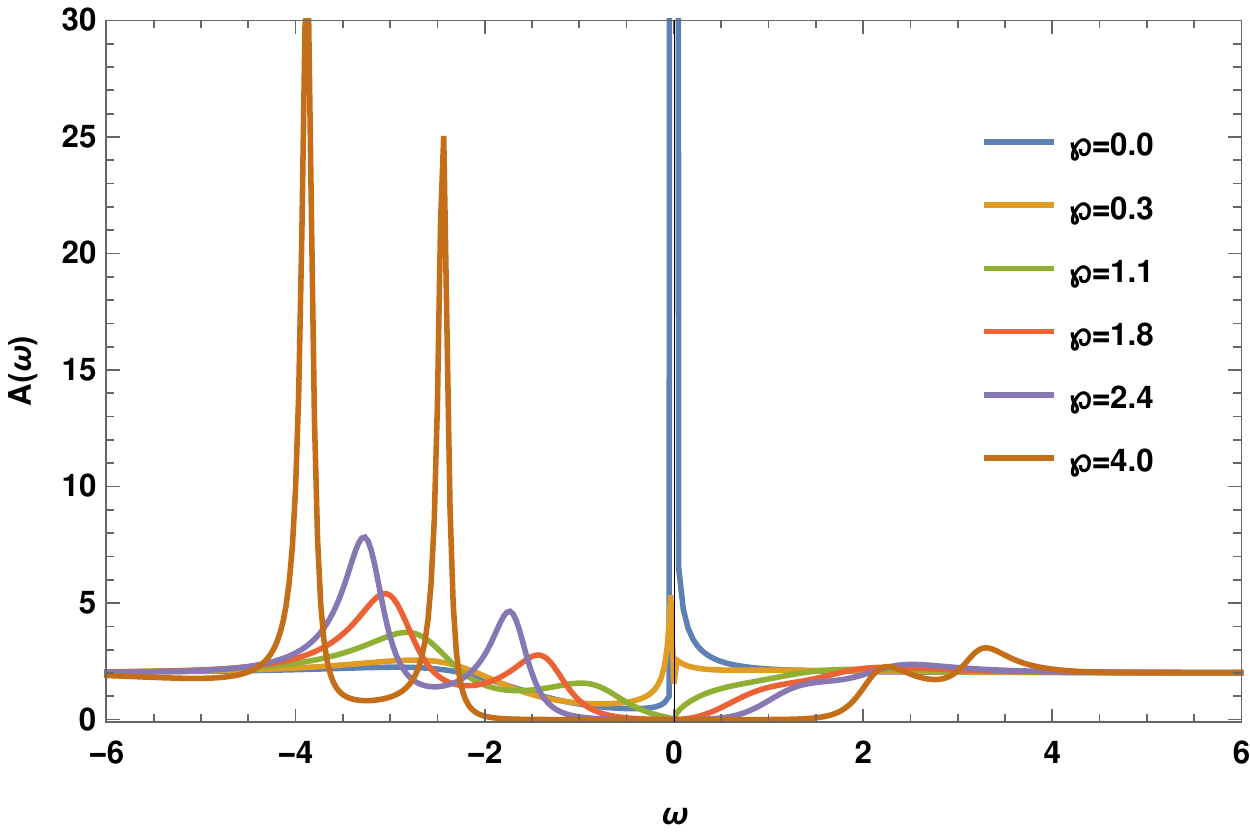}
		\caption{Spectral function $A(\omega)$ at zero temperature for the fermion mass $m=0$ and charge  $q_0=1$. We fix $n=1$, and with $h=0.38$ . }
		\label{fig2}
	\end{center}
\end{figure}
%

\begin{figure}[htbp]
	\begin{center}
		\includegraphics[width=0.75\columnwidth]{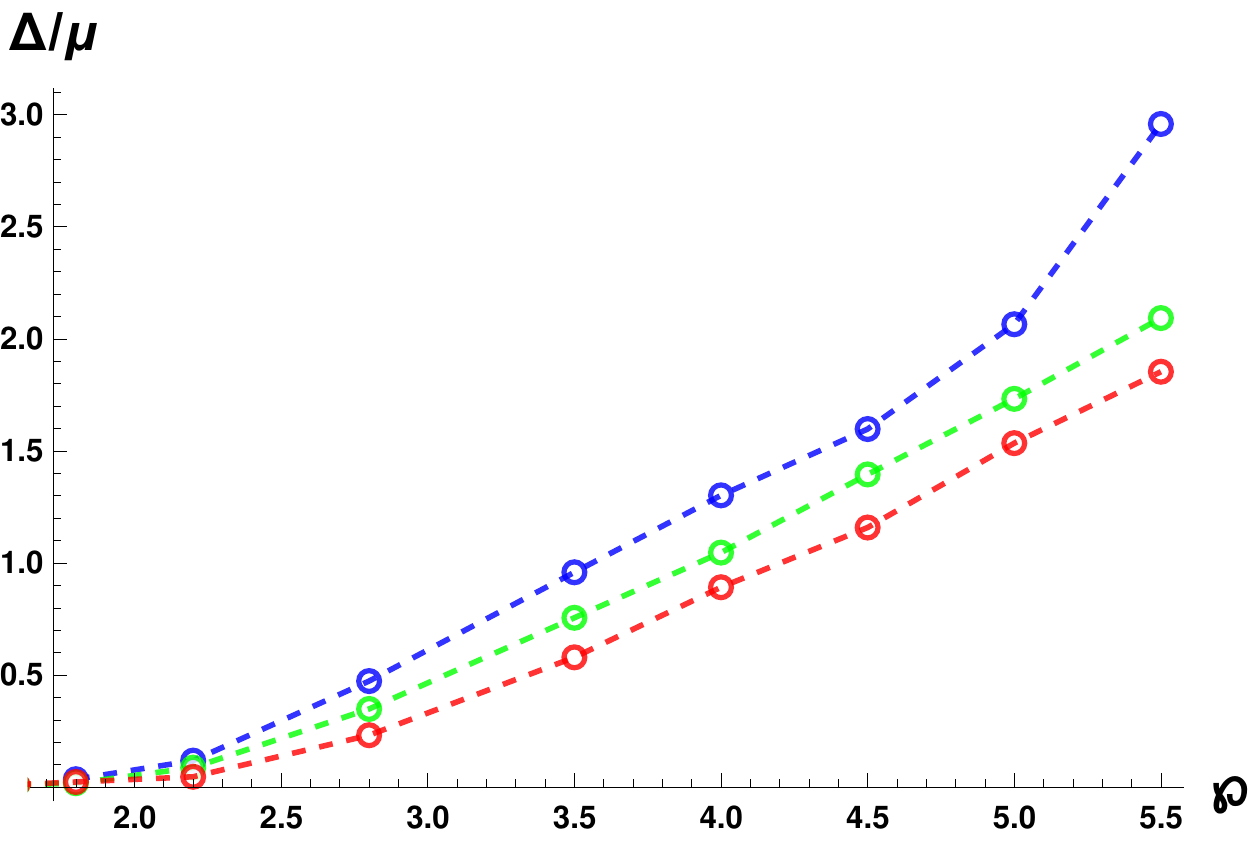}
		\caption{Zero temperature Gap $\Delta$ near $\omega=0$ plotted vs coupling $\wp$ for $n=1$(blue), $n=2$(green) and $n=3$(red). Here magnetic field $h$ is taken from Figure \ref{fig0}, which corresponds to the location of the quasiparticle peak and the fermion mass $m=0$ and charge  $q_0=1.$}
		\label{fig3}
	\end{center}
\end{figure}

\nt In the above discussion, we have studied the quasiparticle decay width for different $n$ levels  for fixed magnetic field $h$. Now  in Figure \ref{fig01} and \ref{fig2}, we observed two interesting  features. Firstly, there is a transfer of the spectral weight. Secondly we  see that a dynamical gap is induced because of the coupling parameter $\wp$. In both Im$G_{22}$ and the full spectral function $A(\omega)$, the appearance of the gap at $\omega=0$ is visible on increasing the value of $\wp$ with a fixed $h=0.38$. The fermionic spectra in this case is way more complicated than that at zero coupling. On comparing the plots for various strengths of $\wp$ in Figure \ref{fig2}, for a fixed $h$ there is a splitting in the spectrum into multiple peaks when one increases $\wp$. This is a new feature that does not appear in a non-dyonic case. This splitting is unique and can only be seen in the presence of a magnetic field with the dipole coupling term introduced in the Fermionic action. From the condensed matter perspective, for low temperatures this was found to be originated from a quadrupole interaction, which implies that there is a change in the local charge distribution \cite{Lu}. On the other hand, on increasing the coupling constant from $\wp=0.2,\;\text{to}\;4$ as shown in Figure \ref{fig2}, apart from splitting in a region near $\omega=0$ when parameter $\wp$ is increased, we also see a dynamical gap as pointed out earlier. We want to study the phenomena of opening of the gap as a function of the dipole parameter $\wp$. In Figure \ref{fig3}, we showed for three $n$ values, the zero temperature gap  opening up at some critical value of $\wp\approx 1.8$ and the gap increases with increasing $\wp$. So far there are few main differences between the fermionic spectral properties  in a dyonic and a non dyonic background. In the case of a dyonic black hole, apart from a spectral weight transfer and opening of a Mott gap, we see the splitting of spectrum into multiple peaks, which is a signature of the magnetic field effects. 
\section{Conclusions}
\label{conclusions}
\nt In this paper we have studied the holographic fermionic spectral function, taking into account the effects of an external magnetic field along with a non-minimal coupling. We also find the Landau's levels arising from the quantization of electron orbits. We observed that the results are different from the case of zero magnetic field. From our results, we have observed some interesting phenomena like the disappearance of the quasi particles for large magnetic field, the splitting of spectral function for large coupling constant, which we have not seen in a non magnetic systems. In addition, we also see a spectral weight transfer from upper band to a lower band by inducing a gap at the Fermi level for different Landau levels $n$. Further, we studied the decay width of the quasiparticles, which shown to have a unique scaling exponent for all $n$ levels. Finally, we studied the variation of the zero temperature gap parameter $\Delta$ versus coupling parameter $\wp$ for the levels $n=1\,2$ and $3$ respectively.

In this paper, we have left out some important aspects, such as transport properties. One can further extend our study with this type of coupling to examine other interesting properties such as the Quantum Hall Effect \cite{Davis:2008nv,KeskiVakkuri:2008eb}, computation of resistivity and conductivity in this system. It would be also interesting to study other intriguing phenomena induced by magnetic field, such as the de-Haas-van Alphen effect, but for now we leave these analyzes for future exploration.


\begin{thebibliography}{10}
	
	
	\bibitem{Anderson:1990}
	P.~W. Anderson.
	\newblock {\em Phys. Rev. Lett.}, 64:1839, 1990.
	
	\bibitem{Varma:1989}
	Varma, Littlewood, Schmitt-Rink, Abrahams, and Ruckenstein.
	\newblock {\em Phys. Rev. Lett.}, 63:1996, 1989.
	
	\bibitem{Maldacena:1997re}
	Juan~Martin Maldacena.
	\newblock {The large N limit of superconformal field theories and
		supergravity}.
	\newblock {\em Adv. Theor. Math. Phys.}, 2:231--252, 1998, hep-th/9711200.
	
	\bibitem{Liu:2009dm} 
	H.~Liu, J.~McGreevy and D.~Vegh,
	\emph{Non-Fermi liquids from holography},
	\emph{Phys. Rev. D} {\bf 83} (2011) 065029.
	\bibitem{Lee:2008xf}
	Sung-Sik Lee.
	\newblock {A Non-Fermi Liquid from a Charged Black Hole: A Critical Fermi
		Ball}.
	\newblock 2008, 0809.3402.
	
	\bibitem{Cubrovic:2009ye}
	Mihailo Cubrovic, Jan Zaanen, and Koenraad Schalm.
	\newblock {Fermions and the AdS/CFT correspondence: quantum phase transitions
		and the emergent Fermi-liquid}.
	\newblock 2009, 0904.1993.
	
	
	\bibitem{Hartnoll:2008vx} 
	S.~A.~Hartnoll, C.~P.~Herzog and G.~T.~Horowitz,
	\emph{Building a Holographic Superconductor}, \emph{Phys. Rev. Lett.}  {\bf 101} (2008) 031601.
	
	
	\bibitem{Hartnoll:2009sz}
	Sean~A. Hartnoll.
	\newblock {Lectures on holographic methods for condensed matter physics}.
	\newblock 2009, 0903.3246.
	
	
	
	
	\bibitem{Herzog:2009xv}
	Christopher~P. Herzog.
	\newblock {Lectures on Holographic Superfluidity and Superconductivity}.
	\newblock 2009, 0904.1975.
	
	\bibitem{Chakrabarti:2019gow}
	S.~Chakrabarti, D.~Maity and W.~Wahlang,
	``Probing the Holographic Fermi Arc with scalar field: Numerical and analytical study,''
	JHEP \textbf{07}, 037 (2019)
	doi:10.1007/JHEP07(2019)037
	
	\bibitem{Faulkner:2011tm}
	T.~Faulkner, N.~Iqbal, H.~Liu, J.~McGreevy and D.~Vegh,
	``Holographic non-Fermi liquid fixed points,''
	Phil. Trans. Roy. Soc. \textbf{A 369} (2011), 1640
	doi:10.1098/rsta.2010.0354
	
	\bibitem{Faulkner:2009wj} 
	T.~Faulkner, H.~Liu, J.~McGreevy and D.~Vegh, \emph{Emergent quantum criticality, Fermi surfaces, and AdS(2)},
	\emph{Phys. Rev. D.}  {\bf 83}  (2011) 125002.

	
	
	
	\bibitem{Andrade:2020hpu}
	T.~Andrade, M.~Baggioli and A.~Krikun,
	``Phase relaxation and pattern formation in holographic gapless charge density waves,''
	JHEP \textbf{03} (2021), 292
	doi:10.1007/JHEP03(2021)292
	
	\bibitem{Cremonini:2019fzz}
	S.~Cremonini, L.~Li and J.~Ren,
	``Spectral Weight Suppression and Fermi Arc-like Features with Strong Holographic Lattices,''
	JHEP \textbf{09} (2019), 014
	doi:10.1007/JHEP09(2019)014
	
	\bibitem{Balm:2019dxk}
	F.~Balm, A.~Krikun, A.~Romero-Berm\'udez, K.~Schalm and J.~Zaanen,
	``Isolated zeros destroy Fermi surface in holographic models with a lattice,''
	JHEP \textbf{01} (2020), 151
	doi:10.1007/JHEP01(2020)151
	
	
	\bibitem{Ling:2013aya}
	Y.~Ling, C.~Niu, J.~P.~Wu, Z.~Y.~Xian and H.~b.~Zhang,
	``Holographic Fermionic Liquid with Lattices,''
	JHEP \textbf{07} (2013), 045
	doi:10.1007/JHEP07(2013)045
	
	\bibitem{Ling:2013nxa}
	Y.~Ling, C.~Niu, J.~P.~Wu and Z.~Y.~Xian,
	``Holographic Lattice in Einstein-Maxwell-Dilaton Gravity,''
	JHEP \textbf{11} (2013), 006
	doi:10.1007/JHEP11(2013)006
	
	\bibitem{Andrade:2017ghg}
	T.~Andrade, A.~Krikun, K.~Schalm and J.~Zaanen,
	``Doping the holographic Mott insulator,''
	Nature Phys. \textbf{14} (2018) no.10, 1049-1055
	doi:10.1038/s41567-018-0217-6
	
	\bibitem{Cremonini:2018xgj}
	S.~Cremonini, L.~Li and J.~Ren,
	JHEP \textbf{12} (2018), 080
	doi:10.1007/JHEP12(2018)080
	
	
	\bibitem{PhysRevLett.106.091602}
	M.~Edalati,  Leigh, Robert G. Leigh, and Philip W. Phillips
	\emph{Phys. Rev. Lett.} {\textbf{106}}, 091602 
	
	\bibitem{Vanacore:2014hka}
	G.~Vanacore and P.~W.~Phillips,
	\emph{Phys. Rev. D} \textbf{90} (2014) no.4, 044022
	doi:10.1103/PhysRevD.90.044022
	
	\bibitem{Gubankova:2010rc}
	E.~Gubankova, J.~Brill, M.~Cubrovic, K.~Schalm, P.~Schijven and J.~Zaanen,
	Phys. Rev. D \textbf{84}, 106003 (2011)
	doi:10.1103/PhysRevD.84.106003
	[arXiv:1011.4051 [hep-th]].
	
	\bibitem{Basu:2009qz}
	P.~Basu, J.~He, A.~Mukherjee and H.~H.~Shieh,
	Phys. Rev. D \textbf{82}, 044036 (2010)
	doi:10.1103/PhysRevD.82.044036
	
	\bibitem{Albash:2009wz}
	Tameem Albash and Clifford~V. Johnson.
	\emph {Holographic Aspects of Fermi Liquids in a Background Magnetic Field}.
	\newblock 2009, 0907.5406.
	
		
	
	
	
	\bibitem{Lu}
	Lu, L., Song, M., Liu, W. et al. Magnetism and local symmetry breaking in a Mott insulator with strong spin orbit interactions. {\em{Nat Commun}} {\textbf{8}}, 14407 (2017).
	

	\bibitem{Davis:2008nv}
	Joshua~L. Davis, Per Kraus, and Akhil Shah.
	\newblock {Gravity Dual of a Quantum Hall Plateau Transition}.
	\newblock {\em JHEP}, 11:020, 2008, 0809.1876.
	
	\bibitem{KeskiVakkuri:2008eb}
	Esko Keski-Vakkuri and Per Kraus.
	\newblock {Quantum Hall Effect in AdS/CFT}.
	\newblock {\em JHEP}, 09:130, 2008, 0805.4643.
	
\end{thebibliography}
\end{document}